\documentclass[10pt, conference, compsocconf]{IEEEtran}
\usepackage{cite}
\usepackage{amsmath,amssymb,amsfonts}
\usepackage{algorithmic}
\usepackage{graphicx}
\usepackage{textcomp}
\usepackage{xcolor}

\usepackage{multirow}
\usepackage{algorithm}
\usepackage{booktabs}
\usepackage{graphicx}

\ifCLASSINFOpdf
\else
\fi
\begin{document}
%
\title{TADO: Time-varying Attention with Dual-Optimizer Model}
\author{\IEEEauthorblockN{Yuexin Wu\IEEEauthorrefmark{2},Tianyu Gao\IEEEauthorrefmark{3}, Sihao Wang\IEEEauthorrefmark{4}\IEEEauthorrefmark{1}, Zhongmin Xiong\IEEEauthorrefmark{2}\IEEEauthorrefmark{1}\footnotemark}
\IEEEauthorblockA{\IEEEauthorrefmark{2}College of Information Technology, Shanghai Ocean University, Shanghai, China\\
\IEEEauthorrefmark{3}College of Software Engineering, Sichuan University, Chengdu, China\\
\IEEEauthorrefmark{4}Department of Mathematics, Southern Methodist University, Dallas, TX, USA\\
\{yuexinwu008,ericgtyu\}@gmail.com, sihaow@smu.edu, zmxiong@shou.edu.cn}
}

%


\maketitle

\renewcommand{\thefootnote}{\fnsymbol{footnote}}
\footnotetext[1]{Corresponding author.}

\begin{abstract}
The review-based recommender systems are commonly utilized to measure users’ preferences towards different items. In this paper, we focus on addressing three main problems existing in the review-based methods. Firstly, these methods suffer from the class-imbalanced problem where rating levels with lower proportions will be ignored to some extent. Thus, their performance on relatively rare rating levels is unsatisfactory. As the first attempt in this field to address this problem, we propose a flexible dual-optimizer model to gain robustness from both regression loss and classification loss. Secondly, to address the problem caused by the insufficient contextual information extraction ability of word embedding, we first introduce BERT into the review-based method to improve the performance of the semantic analysis. Thirdly, the existing methods ignore the feature information of the time-varying user preferences. Therefore, we propose a time-varying feature extraction module with bidirectional long short-term memory and multi-scale convolutional neural network. Afterward, an interaction component is proposed to further summarize the contextual information of the user-item pairs. To verify the effectiveness of the proposed TADO, we conduct extensive experiments on 23 benchmark datasets selected from Amazon Product Reviews. Compared with several recently proposed state-of-the-art methods, our model obtains significantly gain over ALFM, MPCN, and ANR averagely with 20.98\%, 9.84\%, and 15.46\%, respectively. Further analysis proves the necessity of jointly using the proposed components in TADO.

\end{abstract}

%
\IEEEpeerreviewmaketitle

\section{Introduction}
With the increasing popularity of E-commence, personalized recommendations have extensive applications in many aspects of life and can bring up great commercial beneﬁts. Collaborative filtering~\cite{b1} is widely used in personalized recommender system. 
It is mostly based on matrix decomposition, which decomposes item score into latent vectors that can represent users' preferences and products' features. Collaborative filtering-based techniques perform well when there is sufficient score information but would suffer from data sparsity when they don't have sufficient score information~\cite{b1}. To this end, researchers have proposed many recommender systems based on online reviews, aiming to incorporate valuable information from users' review text into the user modeling and recommendation~\cite{b13,b14,
b15}. 
With the continuous development of deep learning, the recommender system based on review text analysis has also achieved promising results~\cite{b11,b13}. Within a large number of theoretical studies and experiments, work~\cite{b11} uses MLP to fit the interaction function and obtain a reasonable overall performance. Work~\cite{b13} uses a convolutional neural network to scan the textual review document through convolution kernels to extract feature representations of users and items. 
Work~\cite{b14} models the reviews independently and then combine the summarized features later.
Furthermore, researchers suggest that the aspect and its weight should be varied when users score towards different items. Hence, the attention neural network server as a solution to ﬁnd out importance among different aspects between different user-item pairs. Work~\cite{b15} applies the attention neural network to analyze review texts, which enables the model to capture the dynamic and fine-grained interaction information between users and items.

The above models preprocess review texts with word embedding. However, work~\cite{b16} argues that there are two problems with this method. First, in the aspect-level model, due to the diversity of users' expression, common aspects are very sparse in most cases. Secondly, the complexity of word combinations in language has certain limitations in extracting the characteristics of review texts, which is because it can't fully explore the complex semantics. Although using a pre-trained word embedding method and fine-tune the training model can help extract the contextual information of the massive reviews, but we think this method has limited ability in analyzing the deep global semantic. Therefore, we encode review texts with BERT~\cite{b17}, which achieves state-of-the-art performance in many NLP tasks, and our experimental result suggests that it also significantly outperform several aspect-level models with word embedding on most datasets.

Additionally, we propose a dual-optimizer model TADO~\footnote[5]{Source code of TADO: https://github.com/woqingdoua/TADO} to effectively alleviate the class-imbalanced problem in the real-world rating prediction task. We found that this problem can greatly limit the performance of the model in predicting ratings with lower proportion. In the Amazon review dataset, reviews with ratings of 4 and 5 account for about $70\%$ of the total dataset, while reviews with ratings of 1 and 2 account for less than $10\%$. We argue that this problem will make the model unable to learn from the low-proportion classes and prone to give a higher rating as the prediction result. Therefore, we design a flexible dual-optimized model to learn a weight for improving the performance in prediction on class-imbalanced data. 
Different from the traditional model with a single loss function for optimization, our model combines classification and regression loss functions with two optimizers to update their parameters, respectively. 
The classification loss function serves as the rating classifier, and the regression loss function aims at learning a weight allocated to different ratings to adjust bias prediction results. Moreover, we employ residual layers and add extra interaction component before the prediction layer. Finally, based on the attention neural network, we add bidirectional long short-term memory and convolution operations with different kernel sizes to capture users' time-varying information.

To verify the effectiveness of the proposed TADO, we conduct extensive experiments on 23 real-world benchmark datasets, and the results demonstrate that our method achieves the state-of-the-art prediction performance. Moreover, we prove the necessity of jointly using these proposed modules together by conducting an ablation study.



\begin{figure*}[h!]
\centering
{
	\includegraphics[width=1\linewidth]{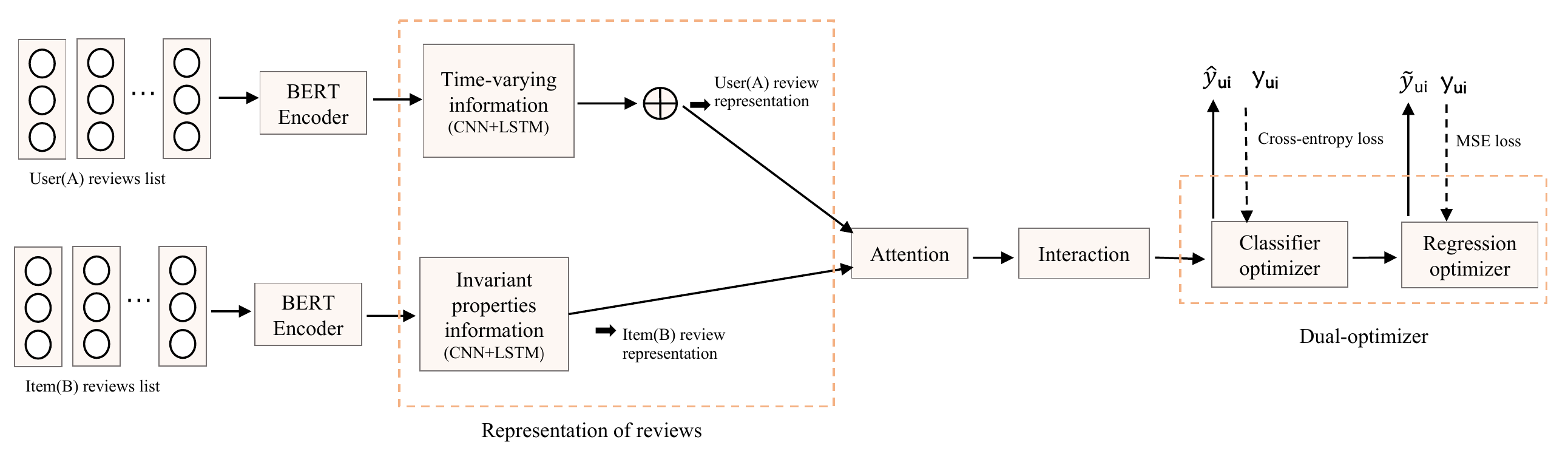}
  \caption{Overall architecture  of the proposed model.}\label{fig:model1}
}
\end{figure*}

\begin{figure*}[h!]
\centering
{
	\includegraphics[width=1\linewidth]{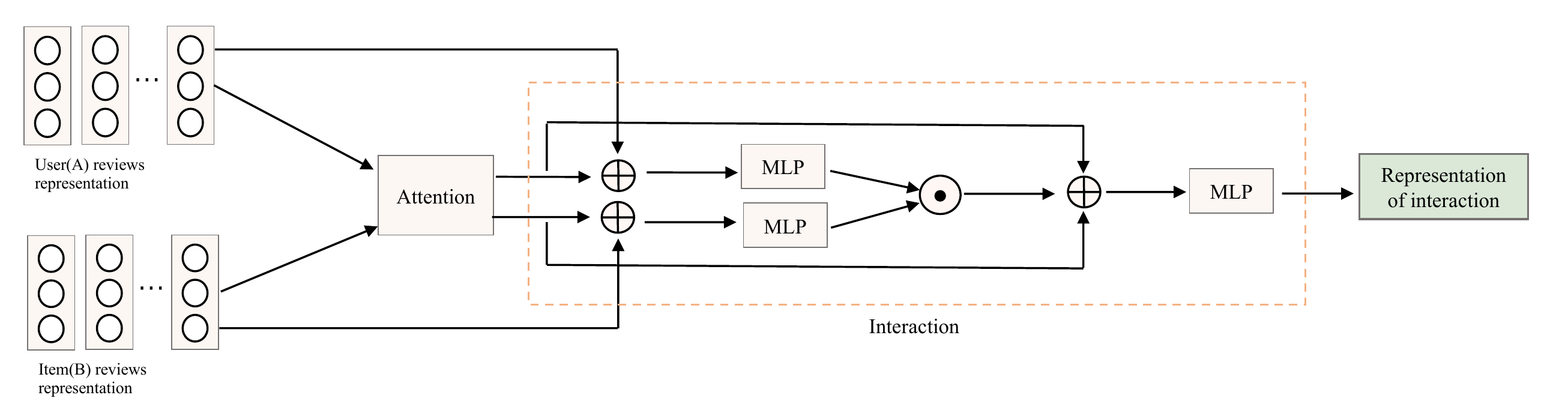}
  \caption{Details of the Interaction Component.}\label{fig:model2}
}
\end{figure*}

\section{Methodology}
In this section, we present the overall architecture of our model and the detail of the interaction component in Figure~\ref{fig:model1} and Figure~\ref{fig:model2}, respectively.

\subsection{Encodding layer}
Our model accepts two input sequences. We represent all users (or items) as a sequence of reviews. $\mathbf{u}=(u_{1},u_{2}...u_{n})$ is a list of users' historical reviews, where $n$ is the maximum number of reviews of different users. Likewise, $\mathbf{i}=(i_{1},i_{2}...i_{k})$ is a list of items' historical reviews, where $k$ is the maximum number of reviews of different items. We keep raw reviews without dropping stop words and encoding them with BERT$_{\mathrm{BASE}}$ model, which contains an encoder with 12 transformer blocks, 12 self-attention heads, and the hidden size of 768. We take the BERT$_{\mathrm{BASE}}$'s final hidden state 768 of the first token \texttt{[CLS]} as the representation of a review. In this layer, all the model's output is a matrix $W \in \mathbb{R}^{d \times |V|}$ where d is the maximum number of reviews and $|V|$ is the hidden size of BERT$_{\mathrm{BASE}}$ model.

\subsection{Feature Representation}
\subsubsection{Users' Feature Representation}
Considering the users' preferences that change over time, we employ bidirectional long short-term memory (Bi-LSTM) to capture time-sequence information of users. Moreover, the multi-scale CNN is utilized to capture long, middle, and short-term preferences with kernel sizes of $1 \times 1$, $3 \times 1$, and $5 \times 1$, respectively. By concatenating their output, we can obtain the representation of the user. Give the input $\mathbf{u}\in \mathbb{R}^{n \times |V|}$, we can obtain the users' feature $\mathbf{f_{u}} \in \mathbb{R}^{r \times |V|}$, where $r$ is the total number of feature vectors obtained from different convolution operations and LSTM networks (e.g. with 3 convolutional kernels and 2 LSTM representations, $r$ equals to $5$ in our method). 
We define the convolution operation as:
\begin{equation}
    \mathrm{Conv}(\mathbf{u})=f\left(\sum_{i, j} \mathbf{u}_{i, j} * k_{i, j}+b\right)
\end{equation}
where $\mathbf{u}_{i, j}$ denotes the sub-region of the input feature vectors with the size of $i \times j$, $f(\cdot)$ denotes the activation function, $k_{i,j}$ is the kernel with the size of $i \times j$, $k$ and $b$ defines the trainable kernel and bias, respectively. Note that the multi-scale CNN is to use different values of $i$ and $j$ to fully extract contextual feature information.
Thus, the user's representation  $\mathbf{f_{u}}\in \mathbb{R}^{r \times |V|}$ can be obtained as follows:
\begin{equation}
    \mathbf{q_{\mathbf{u}}} = \mathrm{stack}[\mathrm{Conv}_{short}(\mathbf{u});\mathrm{Conv}_{middle}(\mathbf{u});\mathrm{Conv}_{long}(\mathbf{u})]
\end{equation}
\begin{equation}
  \mathbf{p_{u}} = \mathrm{BiLSTM}(\mathbf{u})
\end{equation}
\begin{equation}
   \mathbf{f_{u}} = \mathrm{stack}[\mathbf{p_{u}};\mathbf{q_{u}}]
\end{equation}

\subsubsection{Items' Feature Representation}
We assume that items' properties are invariant with time. As such, we use single CNN to extract features of item's reviews as an item's representation. 
Given the input $\mathbf{i}\in \mathbb{R}^{k \times |V|}$, where the features of items $\mathbf{f_{i}} \in \mathbb{R}^{r \times |V|}$ are obtained as follows,

\begin{equation}
  \mathbf{f_{i}} = \mathrm{Conv}(\mathbf{i})
\end{equation}

\subsection{Co-Attention Mechanism}

In this layer, we model the finer-grained interactions to estimate the importance of user and item simultaneously by using the idea of the co-attention mechanism. By using a user's representation $\mathbf{f_{u}} \in \mathbb{R}^{r \times |V|}$ and a item's representation $\mathbf{f_{i}} \in \mathbb{R}^{r \times |V|}$ and feed them into different multilayer perceptrons (MLP), we can obtain the user importance matrix $M_{u}$ and the item importance matrix $M_{i}$ as follows:
\begin{equation}
\left(\begin{array}{c}
\mathbf{Q} \\
\mathbf{K} 
\end{array}\right)=\mathbf{W}\left(\begin{array}{c}
\mathbf{f_{u}} \\
\mathbf{f_{i}}
\end{array}\right) + \mathbf{b}\\
\end{equation}
\begin{equation}
    \mathbf{M} = tanh(\mathbf{K} \mathbf{{Q}}^{\mathrm{T}})
\end{equation}
where $\mathbf{W} \in \mathbb{R}^{r \times r}$ and $\mathbf{b} \in \mathbb{R}^{r \times |V|}$ are the trainable matrix and bias, respectively, $\mathbf{Q} \in \mathbb{R}^{r \times |V|}$ and $\mathbf{K} \in \mathbb{R}^{r \times |V|}$ are the representation of the user and item, respectively. $tanh(\cdot)$ is activate function and $tanh(x) = \frac{e^{z}-e^{-z}}{e^{z}+e^{-z}}$.
Then, we employ co-attention to calculate user importance $\mathbf{M_{u}} \in \mathbb{R}^{r \times r}$,  and vice versa, $\mathbf{M_{i}} \in \mathbb{R}^{r \times r}$ is item importance, as follows:
\begin{equation}
     M_{u,j}=\frac{\exp \left(M_{j}\right)}{\sum_{k=1}^{r} \exp \left(M_{k}\right)}, M_{u,j} \in(0,1)
\end{equation}
\begin{equation}
    M_{i,j}=\frac{\exp \left(M_{j}^{\mathrm{T}}\right)}{\sum_{k=1}^{r} \exp \left(M_{k}^{\mathrm{T}}\right)}, M_{u,j} \in(0,1)
\end{equation}
  
Next, we calculate the final user's representation $\mathbf{Z_{u}}$ and item's representation $\mathbf{Z_{i}}$ for specific user-item pair with the combination of attention $\mathbf{M_u}$ and $\mathbf{M_{i}}$, as follows:
\begin{equation}
  \mathbf{Z_{u}} = \mathbf{M_{u}Q} \hspace{0.5cm} \mathbf{Z_{i}} = \mathbf{M_{i}K}
\end{equation}
where $\mathbf{Z_{u}},\mathbf{Z_{i}} \in \mathbb{R}^{r \times|V|}.$

\subsection{Interaction Component}

The empirical results suggest that an additional prediction layer is beneficial for improving the model's performance. In this layer, we add extra interaction component before the prediction layer by using the residual network. Thus, we can obtain the interaction vector $\mathbf{z}$ as follows:
\begin{equation}
   \mathbf{S_{ui}} = \mathrm{MLP}(\mathrm{concat}[\mathbf{f_{u}};\mathbf{Z_{u}}]) \odot \mathrm{MLP}(\mathrm{concat}[\mathbf{f_{i}};\mathbf{Z_{i}}])
\end{equation}
\begin{equation}
  \mathbf{z} = \mathrm{MLP}(\mathrm{concat}[\mathbf{Z_{u}};\mathbf{Z_{i}};\mathbf{S_{ui}}])
\end{equation}
where $\mathbf{z} \in \mathbb{R}^{1 \times C}$, $C$ is the number of different rating levels. 

\subsection{Prediction}
In this layer, we design a network with dual-optimizer to learn the parameters of the classifier and allocate weight to each rating level.
\subsubsection{Classifier Prediction Layer}
The classifier prediction layer is composed of MLP and softmax function, which converts the input vector into a probability distribution of rating. Given an interaction vector $\mathbf{z}$ between the user and item, the output is obtained as the probability of different rating $p(c)$ and $c \in [1,C]$:
\begin{equation}
  p(c) = \frac{\exp{({z}_{c}})}{\sum_{j=1}^{K} exp({{z}_{j}})}
\end{equation}
where $\sum_{c=1}^{C} p(c)=1$. Here, we denote the distribution of prediction score as $\mathbf{\hat{y}_{ui}} \in \mathbb{R}^{1 \times C}$ and each $\hat{y}_{ui} \in \mathbf{\hat{y}_{ui}}$ equals to $p(c)$.

\subsubsection{Rating Prediction Layer}\label{section:rating_prediction_layer}
Rating prediction layer is a regression model, which has a learnable weight $W \in \mathbb{R}^{1 \times C} $. Finally, we project a value within the target rating range of $[1,C]$ and obtain the predicted score $\widetilde{y}_{ui}$ as follows:
\begin{equation}
  \widetilde{y}_{ui} = 1+\frac{C-1}{1+\exp ( {\sum -W \mathbf{\hat{y}_{ui}}})}
\end{equation}

\subsection{Optimizing \& Learning Strategy}
\subsubsection{Optimizing}
Given a probability distribution of rating $\mathbf{\hat{y}_{ui}} \in \mathbb{R}^{1 \times C}$, we adopt the cross-entropy loss function to optimize our classifier as follows,
\begin{equation}
  \ell_1(\theta_{1}) = \frac{1}{|O|} \sum_{(u, i) \in O} \sum_{l=1}^{C} y_{ij,l} \log \hat{y}_{ij,l}
\end{equation}
where $|O|$ denotes the instance set of user-item pairs. For all $l \neq y_{ui}$,  $y_{ij,l} = 0$ and $y_{ij,l} = 1$ when $l$ equals to ground-truth $y_{ui}$.

Furthermore, we optimize the weight matrix $W$ mentioned in section \ref{section:rating_prediction_layer} by an MSE loss:
\begin{equation}
  \ell_2(\theta_{2}) = \frac{1}{|O|} \sum_{(u, i) \in O}(\widetilde{y}_{ui} - y_{ui})^{2}
\end{equation}

The learning process of weights can be viewed as a regression problem and use the backpropagation technique with the Mean Squared Error (MSE) as the loss function.

\subsubsection{Learning Strategy}
In our model, classification and regression parameters take turns updating in circulation. The learning strategy is shown in Algorithm \ref{algorithm1}. Firstly, we update the classifier's parameter according to the back-propagation of cross-entropy loss. 
It is classifier task in this stage.
Based on the updated classification parameters, we calculate the final rating $\widetilde y_{ui}$ and MSE by the complete model of TADO. According to the result of the back-propagation of MSE, we can update our regression parameter $\theta_{2}$, at this moment, the classification parameters remain unchanged.

\section{Experiments}
In this section, we present the datasets, the comparison baseline models, our experimental setup, and the empirical evaluation. Our experiments are designed to quantify the overall performance and the necessity of jointly using several proposed components.

\subsection{Datasets}
\begin{table*} [h!]
  \caption{Performance comparison on 23 benchmark datasets using the mean squared error. (The best result for each dataset is indicated in bold).}
  \label{tb:comparison}
		\centering
		\resizebox{0.98\textwidth}{!}{
		\begin{tabular}{l c c c c c c c}
			\toprule
			\multirow{2}*{Dataset} & ALFM\cite{b20} & MPCN\& ANR\cite{b15} & TADO(ours) & \multicolumn{3}{c}{\textbf{Improvements\%)}} \\
      ~ & (a) & (b) & (c) & (d) & (d)vs.(a) & (d)vs.(b) & (d)vs.(c) \\
			\midrule
			Automotive & 1.257 & 0.861 & 1.188 & \textbf{0.803}&36.10&6.71&32.39 \\
			App for Android  & 1.555 & 1.494 & \textbf{1.412} & 1.526& 1.86&-2.15&-8.08\\
			Baby & 1.359 & 1.304 & 1.258 &
      \textbf{1.191} & 12.38 & 8.68 & 5.34 \\
			Beauty & 1.466 & 1.386 & 1.387 & \textbf{1.30} &11.32&6.20&6.27\\
			Clothing, Shoes \& Jewelry  & 1.316 & \textbf{1.187} & 1.266 & 1.194 & 9.30 & -0.56 & 5.27\\
			Cell phones \& Accessories & 1.787 & 1.413 & 1.689 & \textbf{1.297} & 27.41 & 8.19 &23.19 \\
			CDs \&  Vinyl & 0.956 & 1.005 & 0.914 & \textbf{0.791} & 17.23 & 21.27 & 13.43 \\
			Digital Music & 0.725 & 0.970 & \textbf{0.688}& 0.839 & -15.68 &13.54 &-21.90 \\
      Electronics &1.563&1.350&1.445&\textbf{1.112}&28.84&17.61&23.02\\
      Grocery \& Gourmet Food &1.284&1.125&1.187&\textbf{0.993}&22.69&11.77&16.38\\
      Health \& Personal Care	&1.466	&1.238	&1.356	&\textbf{1.062}		&27.57	&14.23	&21.69\\
      Home \& Kitchen	&1.443	&1.22	&1.317	&\textbf{1.016}	&29.58	&16.71	&22.84\\
      Instant Video &1.075 &\textbf{0.997} &1.009 &1.028&4.41	&-3.07&-1.84\\
      Kindle Store	&0.870&0.775&0.834	&\textbf{0.664}&23.63&14.27&20.33\\
      Musical Instrument	&1.072	&0.923	&1.034	&\textbf{0.829}		&22.67&10.18	&19.82\\
      Movies \& TV	&1.193	&1.144	&1.112	&\textbf{0.899}	&24.63	&21.40	&19.14\\
      Office Products	&1.474&0.779&1.337&\textbf{0.763}&48.21&2.05&	42.90\\
      Pet Supplies	&1.485	&1.328	&1.377	& \textbf{1.253}&15.61&	5.63	&8.99 \\
      Patio,Lawn \& Garden	&1.510&1.011&1.403&\textbf{1.021}&32.37&	-1.01&27.21\\
      Sport \& Outdoor	&1.221	&0.98	&1.137	&\textbf{0.917}		&24.92&6.45&19.37\\
      Tool \& Home Improvement	&1.384	&1.22	&1.230	&\textbf{0.967}		&30.14	&20.75	&21.39\\
      Toys \& Games	&1.131	&0.973	&1.075	&\textbf{0.823}&27.26&	15.44&23.47\\
      Video Games	&1.383	&1.257	&1.292	&\textbf{1.187}	&14.15&	5.54&	8.10\\

			\bottomrule
		\end{tabular}}
	\end{table*}

In our experiments, we use 23 publicly accessible real-world datasets from Amazon Product Reviews~\footnote{\label{amazon}http://jmcauley.ucsd.edu/data/amazon/}.
These datasets are selected to cover different domains to prove the effectiveness of our proposed method.
The ablation study and some further analyses are conducted on several selected datasets. We utilize a time-based split to partition each dataset into training and testing sets using the ratio (80:20). For Amazon datasets, they are preprocessed in a 5-core fashion (i.e., each user and item have at least 5 reviews to be included). We build users' and items' representation by using each review without interaction.
\begin{algorithm}[!ht]
    \caption{Learning Strategy of the Dual-Optimizer}
    \label{algorithm1}
    \begin{algorithmic}[1]
    \REQUIRE user review set $\mathbf{u}$, item review set $\mathbf{i}$, classification parameters $\theta_{1}$, regression parameters $\theta_{2}$, the times of interaction $T$ and the size of each batch $B$\\
    \ENSURE the prediction rating $\tilde{y}_{\mathrm{ui}}$\\
      \STATE for $t \leftarrow 1$ do
      \STATE \quad  sample a batch ($\mathbf{u},\mathbf{i},y_{ui}$)
      \STATE \quad $\hat{y}_{ui}\gets \mathrm{Classifier\hspace{0.1cm}Prediction\hspace{0.1cm}Layers}(\mathbf{u}, \mathbf{i}, \theta_{1})$
      \STATE \quad $ \ell_1(\theta_{1}) = \frac{1}{|O|} \sum_{(u, i) \in O} \sum_{l=1}^{c} y_{ij,l} \log \hat{y}_{ij,l}$
      \STATE \quad update parameters $\theta_{1} \to
       \theta_{1_{new}}$
      \STATE \quad $\widetilde{y}_{ui}\gets \mathrm{Rating\hspace{0.1cm} Prediction\hspace{0.1cm} Layers}(\mathbf{u}, \mathbf{i},\theta_{1_{new}},\theta_{2})$
      \STATE  \quad $\ell_2(\theta_{2}) = \frac{1}{|O|} \sum_{(u, i) \in O}(\widetilde{y}_{ui} - y_{ui})^{2}$
      \STATE \quad remain the classification parameters $\theta_{1}$  unchanged, and update the
      regression parameters $\theta_{2} \to \theta_{2_{new}}$
      \RETURN $\theta_{1_{new}}$, \hspace{0.05cm} $\theta_{2_{new}}$
    \end{algorithmic}
\end{algorithm}
\subsection{Compared Methods}

We compare our proposed method with three state-of-the-art baseline methods that utilize review information to improve the overall recommendation performance.

\begin{itemize}
 \item \textbf{Aspect-aware Latent Factor Model (ALFM)}~\cite{b20}. ALFM is an aspect-based recommender system that does not rely on external sentiment analysis tools, and is designed as an aspect-aware topic model (ATM) to represent each aspect as a distribution over latent topics.
  \item \textbf{Multi-Pointer Co-Attention Networks for Recommendation (MPCN)}~\cite{b14}. MPCN is a multi-hierarchical model with the multi-pointer to extract important reviews and words from users' and items' reviews dynamically.
  \item \textbf{Aspect-based Neural Recommender (ANR)}~\cite{b15}. ANR is a state-of-the-art end-to-end aspect-based neural approach. ANR performs aspect-based representation learning for both users and items via an attention-based component, and models the multi-faceted process behind how users rate items by estimating the aspect-level user and item importance based on the neural co-attention mechanism.
\end{itemize}

\subsection{Experimental Setup}\label{section:exp_setup}

First, all reviews are encoded by using BERT$_{\mathrm{BASE}}$ model and we only retain \texttt{[CLS]} vector as the representation of a review. We use the cross-entropy loss to train our classifier and the standard Mean Squared Error (MSE) as the evaluation metric of the learning-weight model. 
All models are trained with Adam optimizer~\cite{b38}. We set classifier optimizer with an learning rate of $4 \times 10^{-4}$ and regression optimizer (learning-weight optimizer) of $ 10^{-3}$. our model is trained on all datasets for ten epochs and report the results when MSE on the train sets reaches the lowest. We regularized models with a dropout rate of 0.2, a fixed L2 regularization of $10^{-3}$ in the classifier optimizer, and a fixed L2 regularization of 0 in the regression optimizer. Note that the dropout is applied only before the fully-connected layer of the classifier. We initialize parameters of classifier as $(-2,-1,0,1,2)$ to make it converge quickly. Randomly initialize the parameters can obtain similar or even better performance but it needs more epochs (at around 15 epochs) to converge. We use the Wilcoxon signed-rank test method to verify the signiﬁcance of differences between TADO and other methods. 


\section{Results and Analysis}

\subsection{Experimental Results}

We show the comparison results on all 23 benchmark datasets in Table~\ref{tb:comparison}. It can be observed that our model significantly outperforms ALFM, MPCN, and ANR, with an average improvement of 20.98\%, 9.84\%, and 15.46\%, respectively. For the general standard of Wilcoxon signed-rank test (i.e., $P-value<0.05$), TADO achieves a signiﬁcantly better performance than other approaches.

According to Table~\ref{tb:comparison}, our model achieves improvement in most cases. Compared to MPCN, the relative improvement is encouraging with gains of over 20\% on Tool Home 20.75\%, CDs Vinyl 21.27\%, Movie and TV 21.4\%, respectively. On Musical Instrument, Beauty, Baby, Toy and Games, Grocery Food, Kindle Store, Kitchen, and Electric, a total of 8 datasets, our method obtains a substantial improvement between 10\% and 20\%. Compared to ANR, the best improvement is 42.9\% (on Office Products dataset).
To analyze the reason why our method under-performs on several dataset, we remove the interactive review between user and item in the test stage, in which the preprocessing is different from MPCN and ANR. When we adopt the same preprocessing method as MPCN and ANR, our model gain higher performance across all datasets than MPCN. However, Our improvement is unstable ranging from -15.68\% to 48.21\% when compare with ALFM and ranging from 42.9\% to -21.9\% compared to ANR. 
\begin{table*} [h!]
 \caption{The Comparison Results of the Ablation Study on Five Datasets (Using MSE).}
  \label{tb:ablation}
		\centering
		\begin{tabular}{l c c c c c c }
			\toprule
      \multirow{2}*{Setup}&\multirow{2}*{Video Games}	&\multirow{2}*{Toys \& Games}	&\multirow{2}*{Sports \& Outdoors}	&\multirow{2}*{Beauty}	&\multirow{2}*{Baby}	&Average\\
      ~ & ~ & ~ & ~ & ~ & ~ & (Improvement\%)\\
			\midrule
			TADO (our model)	&\textbf{1.187}	&\textbf{0.823}	&\textbf{0.917}	&\textbf{1.110}	&1.191	&- \\
      \qquad -LSTM	&1.191	&0.858	&0.952	&1.237	&1.236	&-2.90\\
      \qquad -Interaction	&1.252	&0.896	&0.995	&1.289	&1.250	&-6.86\\
      \qquad -Weight Learning	&1.299	&0.991	&1.042	&1.355	&1.411	&-14.77\\
      \qquad -BERT+GloVe	&1.456	&0.912	&0.959	&1.120	&1.201	&-6.28\\
      \qquad -Regression Optimizer Only	&1.193	&0.824	&0.930	&1.263	&\textbf{1.182}	&-0.88\\
			\bottomrule
		\end{tabular}
	\end{table*}
\subsection{Ablation Study of TADO}
We conduct the ablation experiments on the five selected datasets with different scales and domains. The six ablation variants are defined as follows:
\begin{itemize}
  \item \textbf{$-$LSTM}: The model without using the long short-term memory to extract the time series information from the historical reviews of users.
  \item \textbf{$-$Interaction}: A model without the proposed interaction component. After the attention component, the latent representation vectors of user-item pairs are directly fed into the classifier.
  \item \textbf{$-$Weight Learning}: The model without the proposed weight learning component. 
  Cutting the weight learning component means that the model only learned by minimize the cross-entropy loss.
  \item \textbf{$-$BERT$+$GloVe}: This model prepossesses reviews with word embedding. The dimension of the word embedding layer is set to the same as MPCN. 
  The embedding matrix can be initialized using word vectors that have been pre-trained on GloVe~\footnote[8]{https://nlp.stanford.edu/projects/glove/}. We retain words that appear at least 5 times in the reviews and remove the stopwords.

  \item \textbf{$-$Regression Optimizer Only}: A model that retains the classifier component but remove the calculation of cross-entropy loss (only learns from the MSE loss).
\end{itemize}

Firstly, we can observe the remarkable influential component is the weight-learning component, which is proposed to improve learning ability on class-imbalanced data. Moreover, preprocessing by word embedding instead of using BERT to encode the reviews deteriorate performance on four datasets. However, BERT encoding has a decline of performance on beauty 7.51\% while it has a huge performance improvement on Video Games 22.63\%. Furthermore, our model with word embedding still performs better than MPCN on four datasets. 
Finally, we found that removing the interaction component significantly deteriorates the overall performance, which is up to 6.86\% on average. 
In summary, all the proposed components help to improve prediction performance. Among them, the weight-learning is the most effective component, which obtains the improvement of up to 14.77\% on average. Word embedding and the interaction component show a notable improvement, which is 6.28\% and 6.86\%, respectively. The rest components of our model can also slightly improve the performance of less than 1\%. Therefore, all the proposed components are necessary to obtain the promising prediction results.

\section{Conclusion}

In this paper, we propose a novel review-based model named TADO for the recommendation, which includes time-varying attention mechanism, interaction component, and dual-optimizer. 
Moreover, our work it’s the first attempt in this field to alleviate the biased prediction results caused by the class-imbalanced data.
The results show that our model significantly outperforms several state-of-the-art review-based models on 21 out of 23 benchmark datasets from Amazon. 
Furthermore, we conduct extensive analysis on the effectiveness of each proposed component on five datasets. Experimental results have shown that dual-optimizer achieves significant improvement compared to the single classifier optimizer with 14.77\% and the single regression optimizer with 0.88\%. Moreover, the BERT model and the proposed interaction component achieves improvement up to 6.28\% and 6.86\%, respectively.






%

\end{document}